# The Complexification of Engineering

Carlos E. Maldonado and Nelson A. Gómez Cruz

This paper deals with the arrow of complexification of engineering. We claim that the complexification of engineering consists in (a) that shift throughout which engineering becomes a science; thus it ceases to be a (mere) praxis or profession; (b) becoming a science, engineering can be considered as one of the sciences of complexity. In reality, the complexification of engineering is the process by which engineering can be studied, achieved and understood in terms of knowledge, and not of goods and services any longer. Complex engineered systems and bio-inspired engineering are so far the two expressions of a complex engineering.

## 1. INTRODUCTION

We are currently facing a dynamic process of complexification of engineering sciences. To be sure, such is a proof of vitality and change that, nonetheless, is to be fully understood and explained. That is the aim of this paper.

There was a time when engineering was considered just as a practice on based-knowledge [1]; in other words, as the way through which science acts upon the world. As such, engineering could be considered as part of the little science [2] (The expressions "little science" and "big science" will be used repeatedly here. They are to be taken not literally in the sense of De Solla Price, but as references of a turn or a shift in the evolution of complexity. The basic idea is the move from little science towards big science as a trend to the complexification of engineering. This idea is not present in the book by De Solla Price).

The aim of engineering aim was directed toward the production and control of goods and services. At its best, it was part of the third sector of economics, namely services. Physics and mathematics, and particularly integral and differential calculus, were the very core of engineers as they were striving for exact methods and single solutions for each problem at a time. Classical formal logic and particularly boolean systems and notations have been part and still belong to the normal formation of engineers. Engineers were trying to differentiate themselves as they acted upon nature, the world and society in a variety of subfields; for instance, civil, electronic, environmental, chemical, transport, food or aero spatial engineering to name but a few.

The complexification of engineering is however not to be taken as a diversification of its fields and scopes neither in the use of a range of mathematical, technological and computational tools. Something deeper and from a wider scope and reach is at stake both within engineering and in its relationships with other sciences and disciplines. What is truly going on affects the very nature of science and of nature, eventually.

In this paper we study the trend of engineering to complexity. In so doing we define the relationship between engineering and complexity and we claim that engineering is going through a radical change of its very nature, even though such a trend is far from being general or normal. We will focus not so much in the past of engineering as in the process-to-be. At the end, several conclusions are drawn.

## 2. THE STATE-OF-THE-ART AND THE TREND TOWARD COMPLEXIFICATION OF ENGINEERING

Classical engineering cares about prediction, predictability, transparency, reliability and centralized control. Serious concerns about literally building up the world are at stake throughout these concepts and practices. The engineering sciences including the sciences of materials have serious pragmatic interests in what these concepts are all about, no doubt. And yet, the focus of this paper is not about the pragmatics of engineering as in the conceptual if not the scientific and philosophical stand of engineering.

Within what can be called as classical engineering reversed systems engineering plays a fundamental role in that it concentrates in aspects as reuse and on the analysis of existing problems. A number of companies and countries have been growing and developing focusing mainly in reversed systems engineering. What is true is that both classical and reversed systems engineering are well set up as a work and research on intelligent solutions. In general, an intelligent solution in engineering is the one that is based on approximate models and on a variety of heuristic techniques such as tabu search, simulated annealing and stochastic hill climbing, and the like. The methods of modern heuristics include fuzzy systems, neural networks, genetic algorithms and multi-agent systems [3, 4].

More recently, a new conceptual field has been growing that focuses on systems or software that supports engineering activities. The importance of computing, modeling and simulation and even more widely the significance of microelectronic systems and components have triggered the importance of meta-engineering [5].

Meta-engineering has been conceived as a sort of "engineering engineering" with the help of software and/or systems that support all daily activities, choices and plans proposed or carried out by engineers. The focus here is placed on design, and design is conceived as the most fundamental part of engineering at large.

Being as it might be, the present state of engineering – both as a practice as in its conceptual or theoretical foundations, the engineering sciences deal with great success with simple and complicated systems. A simple system is simply one that can be fully grasped and manipulated in terms of division, fragmentation, segregation. As for a complicated system, it is the one that can be understood and intervened with the help of approaches and tools such as statistics, matrix analyses, vectors, standards, averages, Bell or Gauss curves, and the like. Plainly said, a complicated system is a set of simple systems.

## 3. A TWOFOLD POINT OF BIFURCATION

Engineers have being seeing for a while how its field has been growing encompassing a number of domains pertaining to companies and corporations. If classical engineering arose from its interest in controlling and superseding nature (from the 17[th] Century onwards), the new "nature" so to speak has become the world enterprises at large. By this we do not merely mean industrial engineering or the like, but a new array of literally *engineering* society, economics, finances and management. As a consequence a point of bifurcation is located that opens up classical and conventional engineering onto a highway that gets us apart from the traditional concern based on local research and exact methods and closer to a point where knowledge is to be recognized as the turning point if not the axis for engineering sciences. Several authors speak about enterprise systems engineering [6, 7] but following Jamshidi [8, 9] we prefer to stress on systems of systems engineering (SoSE). Our preference lies in the fact that the concept of "system of systems" is wider and deeper than "enterprise", which can be rightly conceived as an element of SoSE as it is indeed the case according to Jamshidi *et al* [10, 11].

Yet, a second point of bifurcation can be clearly identified. It has to do with the very work throughout which non-linearity is worked out – basically for practical or pragmatic stances, in a linear way. This is a sensitive question both in engineering and science in general.

The linearization of the world in general was the very spirit that gave birth to calculus, namely differential and integral calculus by Leibniz and Newton. To be sure, linearizing non-linearity produced practical outcomes for it permitted intervention, control, manipulation and prediction of phenomena. We were then at the heart of modern science at large.

The linearization of non-linear systems allows us characterizing unconventional engineering (UE), a rather blurry concept, for it stands as a contrast of classical and conventional engineering. Unconventional engineering is rather a negative concept – over against, so to speak, classical and conventional engineering.

## 4. STEPPING INTO ENGINEERING´S NEW TERRITORIES

In spite of the segregation and hyper-specialization of engineering – certainly motivated more by the market than by engineering´s own epistemological or scientific evolution, a solid trend toward integration of systems, both horizontal and vertical can be identified. Such integration of systems not only produces fruitful synergies, but it places coordination and interoperability as the very engine, so to speak, of engineering the world and society at large.

System of systems engineering (SoSE) represents not just a bifurcation within the engineering sciences but, more appropriately a new territory for work and research. While it has being paying a debt to classical and conventional engineering, it heads up towards the recognition of information and computing as vital assets for engineering leaving in the backstage the traditional concern for matter and energy. - A deep and complex question arises here concerning the shift from energy and matter towards, or into, information or, in other words, the relationships between energy, matter and information. We can advance that the three are fundamental for the understanding of open systems. Moreover, information is not a mere "third" term in an equation that from the outside appears as: *bit from it, and from bit qu-bit*. Information is an expression of matter and/or energy, and there is no *causal* relationship among the three but a *synthesis*. Here, however, we cannot go further into these ideas, for they are the subject of a different paper with are currently working on. As a hint for what precedes, see [12, 13].

The integration, horizontal and vertical, of systems implies a crucial turning point in engineering in that the interest is not placed any longer in solving (particular or singular) problems but in working with space of solutions. Hence, a shift is introduced that steps forward from heuristics to metaheuristics [14, 15]. Thus, the turn by which information and computation become crucial is strongly intertwined with metaheuristics. In a future-to-come, we speculate, we wish not too long from now, the interface between engineering and super-Turing computation (or hypercomputation [16]) will be a happy encounter and a normal day-to-day practice and dialogue.

A frontier of SoSE brings us to the recognition of the importance – both theoretical and practical, of inaccuracy, learning, and vagueness. These can be said to be the lower bound of the new territories of engineering sciences that question the rigidity of traditional engineering. As the world introduces new materials and discovers a variety of different approaches engineering becomes less attached to the knowledge introduced from a-far and starts to produce knowledge by itself. As a result, engineering crosses the boundary that separates little science from big science and heads up to the production of knowledge as a vital process in engineering.

The existence of inaccuracies is neither a failure for science and engineering nor a sheer epistemological or cognitive feature. On the contrary, it implies necessarily the importance of learning systems and, thereafter, of learning sciences and tools, i.e. rods.

As such, vagueness is not to be confused with fuzziness nor with blurriness. Properly speaking, a system is said to be vague – or permeated by vagueness when more subtle measurements and uncertainty are considered and not excluded any more.

## 5. WHEN ENGINEERING MAKES OF ITSELF A SCIENCE

In our view, the definite complexification of engineering is introduced thanks to the development of complex engineered systems (CES) [17, 18]. In this frame, the complexification is not to be taken as a complication of engineering sciences, but rather as the non-linearization of engineering, which entails the entrance into, or the transformation of engineering as, big science. The non-linearization of engineering opens up engineering to the domain of the possible leaving behind the consideration that the core of engineering are facticities (*Faktizität*).

One salient aspect of big science is that science – at large, does not have the common preoccupations of normal science which can be summarized as: its social relevance, the process of building bridges with the concerns of the public or the private sector as to the impact of the research, the meaningfulness of basic research vis-à-vis applied or experimental research, budgeting the whole scientific exercise and providing an account of every step and process in order to win social, financial and political acceptability, the worry about the gaps with other sciences and disciplines and the continuous defense of its own status and enterprise, for instance.

In the most pregnant sense, the making of engineering a science consists in the fact that the engineering sciences can identify a series of problems of their own by themselves. More notoriously and surprisingly, the altogether new problems can be safely called as *frontier problems*, by which the big sciences can also be viewed as *frontier sciences*. Some of the properties or attributes of such complexity are: emergence, self-organization, synergy, non-linearity, non-rigid and decentralized control(s), the existence of power laws, adaptation, and free-scale networks, the prevailing role of non-classical logics, among others.

One recent fast development to be taken into account is materials sciences – a concept that crosses or encompasses physics, chemistry, and engineering; material sciences are in fact frontier science. We have come to encounter and develop composite materials, and completely new materials, including ceferene and graphene.

Accordingly, the work and understanding of, and with, durability and reliability have become a central concern in engineering. The challenge remains however concerning the self-repairing of systems a problem on which the best lights and insights come from biology.

In any case, engineering complex systems means a radical turn from physics as paradigm toward biology as frame and reference [19]. The biology that is considered in this kind of engineering is certainly not molecular biology, but the *new biology*, a concept that refers sometimes to Evo-Devo as well as to evolutionary biology, to the concepts and theories that tend to complement as well as to supersede the role of natural selection in the theory of evolution. These are concepts such as the derivation of genetic interaction networks, the role of randomness in genetics, indirect selection, self-organization, and the role of statistics both at the developmental and evolutionary levels.

Uncertainty is a pregnant concept in the sciences of complexity as well as, therefore, in complex engineered systems. It is neither a sheer limit of our cognition nor a gap in reality due to the presence of noise or the lack of some information. On the contrary, uncertainty is a positive feature that is to be translated as the openness of a system. Thus, an open system exhibits uncertainty as a constitutive trait.

Engineering complex systems consists in producing systems capable of adaptation, change and novelty, and not any longer systems defined by stability, predictability, reliability and control [20, 21]. A fundamental shift is produced here, namely leaving behind stability and taking up a quite different and novel concept: robustness. Even though there is no single definition of robustness, there are a variety of case studies showing that engineering is moving forward to a complexification of its own enterprise. Robustness allows engineering sciences to focus on the ability of a system to maintain specified features when facing turbulences, whether internal or external [22]. Among the attributes that define CES the following can be mentioned: resilience, evolvability, adaptation, self-repair, robustness, flexibility, and scalability.

A quite sensible point in engineering complex systems is easily stated and yet very hard to implement or work with. The question has to do with how are we to understand the *desirable complexity* that is to be produced [23, 24]. Here more than anywhere else the interplay and positive loops among information, computation and knowledge become fundamental. And yet, the whole process should not be understood in the classical sense of gathering particular observations, playing with degrees of generalization, and rationalizing via maximization or optimization as the whole question. For another aspect plays a crucial role here, namely intuition bubbles, imagination, logics, *gedanken-experimenten* – which have to do, all of them, with the capacity of risk, play, challenge and innovation on the behalf of the researcher, whether engineer or scientist.

Without hesitation we can say that the science that has studied more carefully emergence and the mechanisms throughout which particular emergent phenomena can be produced –i.e., the *desirable complexity*, is artificial life [25-27]. Moreover, much of the uprising within classical and conventional engineering has been produced by artificial life [28]. As a consequence, a sensitive question becomes the one concerning the possible scales of emergence; that is, how to produce or reach more than one level of emergence in the search of a desired global behavior [29].

The problem however with complex engineered systems is that it lacks a theoretical and technical framework [17, 30]. The technical lack has to do with metaheuristics, and more particularly with multi-objective, hybrid, parallel, adaptive and multilevel metaheuristics [15, 31] as well as with hypercomputation in that it does not have a practical application as yet. It is our contention that such a theoretical framework can be provided by the other great component by which engineering becomes big science, namely bio-inspired engineering [32, 33].

Bio-inspired engineering (BIE) can be said to be more radical as to the taken up of biology as the frame and basis for engineering; indeed, learning from natural systems becomes an increasing concern in the way how engineering makes of itself a science or, to put it the other way round, when science meets technology [17].

It is at this stage where definitely goods and services are not any more the references for engineers, but knowledge. In the interplay between CES and BIE the ongoing trend of complexification of engineering reaches its peak, so far.

It is necessary to stress the fact that biology and nature are not just a metaphor or a cultural frame in the new territories where engineering, the sciences of complexity and computation coincide and agree. Moreover, nature, in the broadest but deepest sense of the word, is taken as the most serious motive for research and work. Ranging from the immunological system to the study of ant colonies; from the formations and dying of stars to the problem concerning apoptosis and the biological clock; from ecosystems and the biosphere to economics, military conflicts and the spread and contagion of diseases; from financial and economic crashes to artificial life, swarm robotics and swarm intelligence – nature is taken, for the first time in history as a truly complex system and hence, in an interdisciplinary if not cross-disciplinary approach. Throughout this, engineering in general –and more particularly CES and BIE are starting to learn how to design, engineer and produce a *second nature*, a *second life* [34], a second opportunity for the entire world. It is exactly in this sense that the sciences of complexity are said to be a scientific revolution, or a bifurcation [35].

All in all, engineering matters today more than ever, for it contains the very way through which we can produce, design, build and create a better world with more and better opportunities. The point here is that engineering is not alone in the possibility of such enterprise, for the very concept of big science is useful as to showing how science and technology are becoming one and the same spirit. And the focus of the unification of science and engineering – Not to mention the arousal of the "third culture" [36], indicates that the complexification of engineering is the very same process by which we are coming closer to life, namely to understanding, getting along with it, and making it always more and more possible. In one word, the complexification of engineering is but a way to express the fact that we scientists, engineers and philosophers, for instance, are coming to a *living technology* [37]. A living technology is the one literally inspired by nature and life and taking up the processes of information and computation the living systems are carrying on continuously. Without being pretentious, we are all reaching an ultimate frontier. Figure 1 summarizes our study:

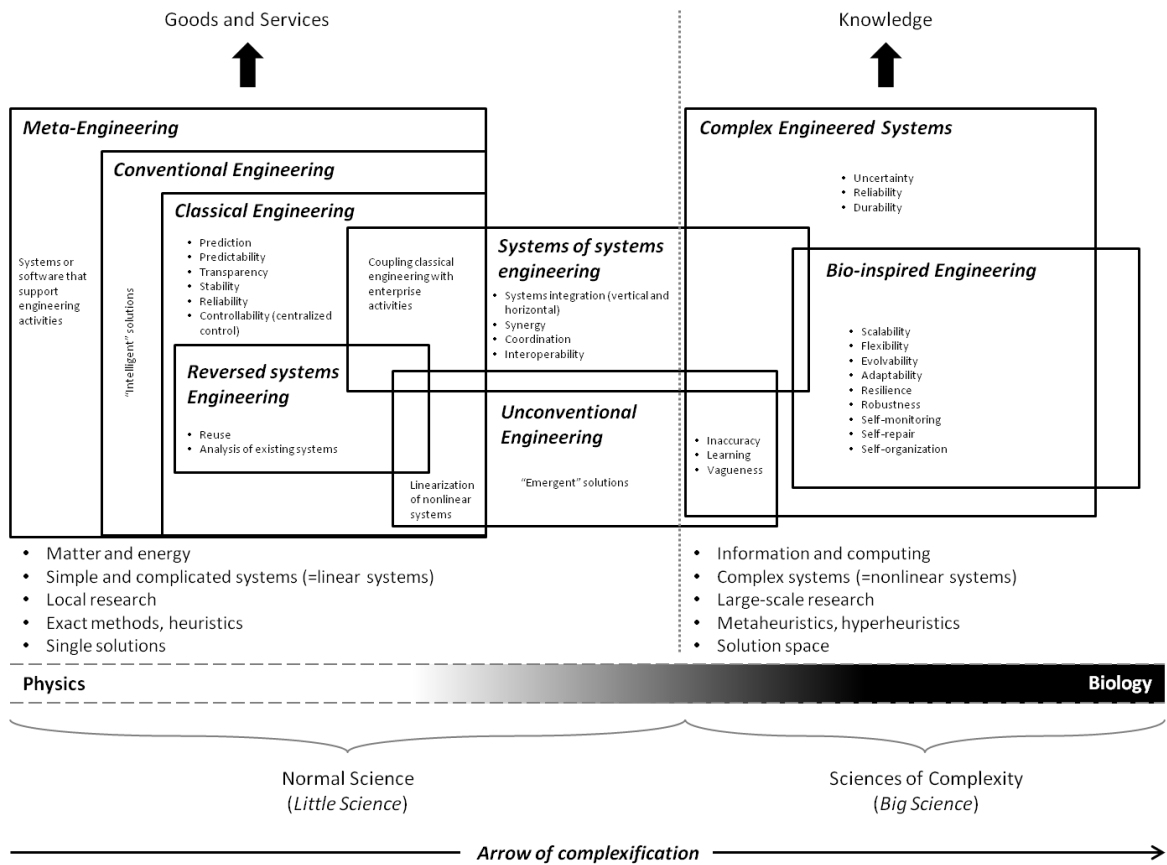

**Figure 1. Complexification of engineering sciences.**

## 6. A CONCLUDING REMARK
To conclude, let us make explicit our concern and interest for engineering. What is the interest in tracing the map of the complexification of engineering? We have a number of reasons, namely: i) understanding what engineering really is like is vital for the comprehension of technology. After all, engineering is today's *techné* and *ars*, and we all need to clarify our ideas concerning technology. Yet it is a very particular *ars* and *techné* in that it has grown into a science in its own. To be sure, that is the subject of philosophy of technology; ii) in the history of mankind the relationships between science and technology have never been so passionate, vital and crucial. In an optimistic view of future our fate depends on knowledge and research, and the engineering sciences play a fundamental role then; iii) acting upon nature and society is a matter of both sensitivity and intelligence. Two ways how we act are science and technology, not to mention art at large. Hence, the question about engineering concerns the way we live and act upon the world. That is a social and cultural matter. All in all, the three reasons just mentioned are of philosophical nature. A fourth reason is to be added which concerns the interplay of non-linearity, computation and information. Well, the engineering sciences are one very serious way that can help us understand, we believe, what life is really made of and what it is about. The last frontier, so to speak, is bio-inspired engineering, precisely. And our interest for life is certainly not a minor one. Our best endeavor is heading up exactly in this direction.